\documentclass[prl,letterpaper,aps,10pt,twocolumn,floatfix,showpacs,superscriptaddress]{revtex4-1} 
\usepackage{graphicx}
\usepackage{amsmath,amssymb,amsfonts}
\usepackage{graphicx,braket,xcolor,subfigure,upgreek}
\usepackage[colorlinks, linkcolor=blue, citecolor=blue, urlcolor=blue, breaklinks=true]{hyperref}
\usepackage{microtype}
\usepackage{bbm}
\usepackage{color}
\usepackage[english]{babel}
\usepackage[toc,page]{appendix}

\usepackage{lipsum}

\begin{document}

\title{Subradiance and Superradiant Long Range Excitation Transport among Quantum Emitter Ensembles in a Waveguide}

\author{Martin Fasser}
\affiliation{Institut f\"ur Theoretische Physik, Universit\"at Innsbruck, Technikerstr. 21a, A-6020 Innsbruck, Austria}
\author{Laurin Ostermann}
\affiliation{Institut f\"ur Theoretische Physik, Universit\"at Innsbruck, Technikerstr. 21a, A-6020 Innsbruck, Austria}
\author{Helmut Ritsch}
\affiliation{Institut f\"ur Theoretische Physik, Universit\"at Innsbruck, Technikerstr. 21a, A-6020 Innsbruck, Austria}
\author{Christoph Hotter}
\email{christoph.hotter@uibk.ac.at}
\affiliation{Institut f\"ur Theoretische Physik, Universit\"at Innsbruck, Technikerstr. 21a, A-6020 Innsbruck, Austria}

\date{\today}

\begin{abstract}
In contrast to free space, in waveguides the dispersive and dissipative dipole-dipole interactions among quantum emitters exhibit a periodic behavior over remarkably long distances. We propose a novel setup exploiting this long-range periodicity in order to create highly excited subradiant states and facilitate fast controlled collective energy transport amongst far-apart ensembles coupled to a waveguide. For sufficiently large ensembles collective superradiant emission into the fiber modes dominates over its free space counterpart. We show that for a large number of emitters a fast transverse coherent pulse can create almost perfect subradiant states with up to $50\%$ excitation. On the other hand, for a coherent excitation of one sub-ensemble above an overall excitation fraction of $50\%$ we find a nearly lossless and fast energy transfer to the ground state sub-ensemble. This transport can be enhanced or suppressed by controlling the positions of the ensembles relative to each other, while it can also be realized with a random position distribution. In the optimally enhanced case this fast transfer appears as superradiant emission with subsequent superabsorption, yet, without a superradiant decay after the absorption. The highly excited subradiant states as well as the superradiant excitation transfer appear as suitable building blocks in applications like active atomic clocks, quantum batteries, quantum information protocols and quantum metrology procedures such as fiber-based Ramsey schemes.
\end{abstract}


\maketitle

\section{Introduction} 
Spatial confinement of electromagnetic radiation field modes increases the atom-field coupling as well as the resulting dipole-dipole interaction between distant quantum emitters as the coupling is proportional to the field amplitude per photon generated by one emitter at the position of another. Extreme cases are optical resonators exhibiting quasi-uniform strong coupling over the entire mode volume~\cite{dovzhenko2018light} or single mode waveguides~\cite{gonzalez2024light} such as optical fibers with weaker but close to infinite range interaction at their center.

As the atoms in a fiber are often distributed over large distances, their direct local interaction via free space is typically very small compared to the collective interaction mediated by the fiber~\cite{ostermann2019superandsub}. Recent advances in fiber fabrication technology, laser cooling and manipulation of cold atomic ensembles allow for the implementation of such collective dynamics at unprecedented atom numbers~\cite{ruks2024coherent,chen2024magnetic,ribezzo2023quantum}. While one typically first observes superradiant collective emission~\cite{dicke1954coherence,gross1982superradiance} into the fiber~\cite{goban2015superradiance,okaba2019superradiance,pennetta2022collective,liedl2024observation}, in this work we find that the opposite case of subradiance~\cite{albrecht2019subradiant, holzinger2022control, zanner2022coherent, tiranov2023collective} as well as superabsorption~\cite{higgins2014superabsorption, yang2021realization, quach2022superabsorption} can be implemented and reveals a plethora of novel physical phenomena and applications, especially in very large ensembles.

In addition to constituting a novel implementation of protected Ramsey spectroscopy~\cite{ostermann2013protected, ostermann2014protected, hotter2023cavity, bohr2024collective} we predict fast efficient long-distance collective energy transfer~\cite{hama2018negative, hama2018relaxation} based on superradiance with subsequent superabsorption. This interplay between subradiant states with strongly reduced emission rates and the superradiant energy transfer can be exploited in quantum memories~\cite{zhao2008long, zhao2008millisecond, lvovsky2009optical, plankensteiner2015selectiv, asenjo2017exponential}, quantum batteries~\cite{ferraro2018high, pirmoradian2019aging, munro2020darkstates, ferioli2021storage, quach2022superabsorption} and in order to remotely prepare perfectly inverted ensembles~\cite{munro2021energymigration}.
\begin{figure}[t]
\includegraphics[width=0.99\columnwidth]{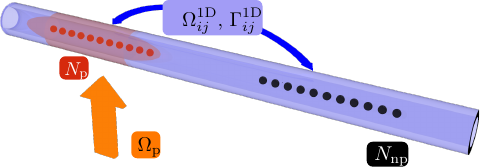}
\caption{\emph{Model.} Two ensembles of atoms coupled to a waveguide mode. Initially, only one ensemble with $N_\mathrm{p}$ atoms is pumped while the other with $N_\mathrm{np}$ atoms remains in the ground state. Below a certain pump threshold an almost perfectly subradiant state is created, while above this threshold a fast superradiant energy transfer from the pumped to the non-pumped ensemble takes place, where the energy is (super-)absorbed and remains in the second ensemble.}
\label{fig:model}
\end{figure}

The manuscript has the following structure: after introducing our general model based on quantum emitter arrays with fiber-mediated effective dipole-dipole interactions we focus on the special case of two far-apart ensembles. In order to treat a large number of emitters and high excitation manifolds we employ cumulant expansion equations up to second order to numerically study generic cases of the collective system dynamics, when only one of two ensembles is coherently pumped. We then analyze the transition from a subradiant collective decay blockade to a superradiant energy transfer between the two ensembles as a function of atom numbers, coupling conditions and initial state.

\section{Model}
We consider $N$ identical two-level emitters coupled to a single optical fiber mode, see fig.~\ref{fig:model}. The coupling rate of each emitter into the fiber is given by $\Gamma^{1\mathrm{D}} = \Gamma \frac{k_0 \mu^2}{\epsilon_0 A} \chi$, which is determined by the free space decay rate $\Gamma$, the effective area $A$, the mode wavenumber $k_0$ and the normalization constant $\chi$ that accounts for the distribution of the electric field into the mode~\cite{ostermann2019superandsub,cardenas2023many}. In the rotating frame of the atomic transition frequency, the system is described by the Hamiltonian 
\begin{equation}
    H= \sum_{i \neq j}^{N} \Omega_{ij}^{1\mathrm{D}} \sigma^+_i \sigma^-_j.
\end{equation}

The waveguide-mediated dipole-dipole coupling rates are
\begin{equation}
    \Omega^{1\mathrm{D}}_{ij}(x_{ij})=\frac{\Gamma^{1\mathrm{D}}}{2} \sin{(k_\mathrm{eff} x_{ij})},
\end{equation}
where $x_{ij}= | x_i - x_j |$ is the distance between two emitters $i$ and $j$, and $k_\mathrm{eff}$ denotes the effective wavenumber of the guided mode. The incoherent part describing the dissipative processes is accounted for by the Lindblad term
\begin{equation}
    \mathcal{L}[\rho]=\frac{1}{2}\sum_{i,j}^N \Gamma^{1\mathrm{D}}_{ij} (2 \sigma^-_i \rho \sigma^+_j-\sigma^+_i \sigma^-_j \rho - \rho \sigma^+_i \sigma^-_j)
\end{equation}
with the collective fiber-mediated decay rates
\begin{equation}
    \Gamma^{1\mathrm{D}}_{ij}(x_{ij})=\Gamma^{1\mathrm{D}} \cos{(k_\mathrm{eff} x_{ij})}.
\end{equation}
We assume sufficiently large distances between the emitters such that we can neglect free space dipole-dipole interaction, which decreases as $1/x$ in the far field. Furthermore, we are only interested in time scales, where the individual free space spontaneous emission of the atoms is not relevant. The open system can thus be described by the following master equation in the Born-Markov approximation~\cite{gardiner2004quantum}
\begin{equation}
    \Dot{\rho}=i[\rho, H]+ \mathcal{L}[\rho].
\end{equation}

Since we aim at considering large systems with up to $10^6$ emitters, where a full quantum simulation is by far out of reach, we will treat this problem in a second-order cumulant expansion \cite{kubo1962generalized, plankensteiner2021quantumcumulantsjl} which turns out to capture the dynamics remarkably well. A code example that automatically derives and numerically solves the corresponding equations is provided in the appendix.

\section{Subradiant States}
\begin{figure*}[tb]
    \centering
    \includegraphics[width=0.99\linewidth]{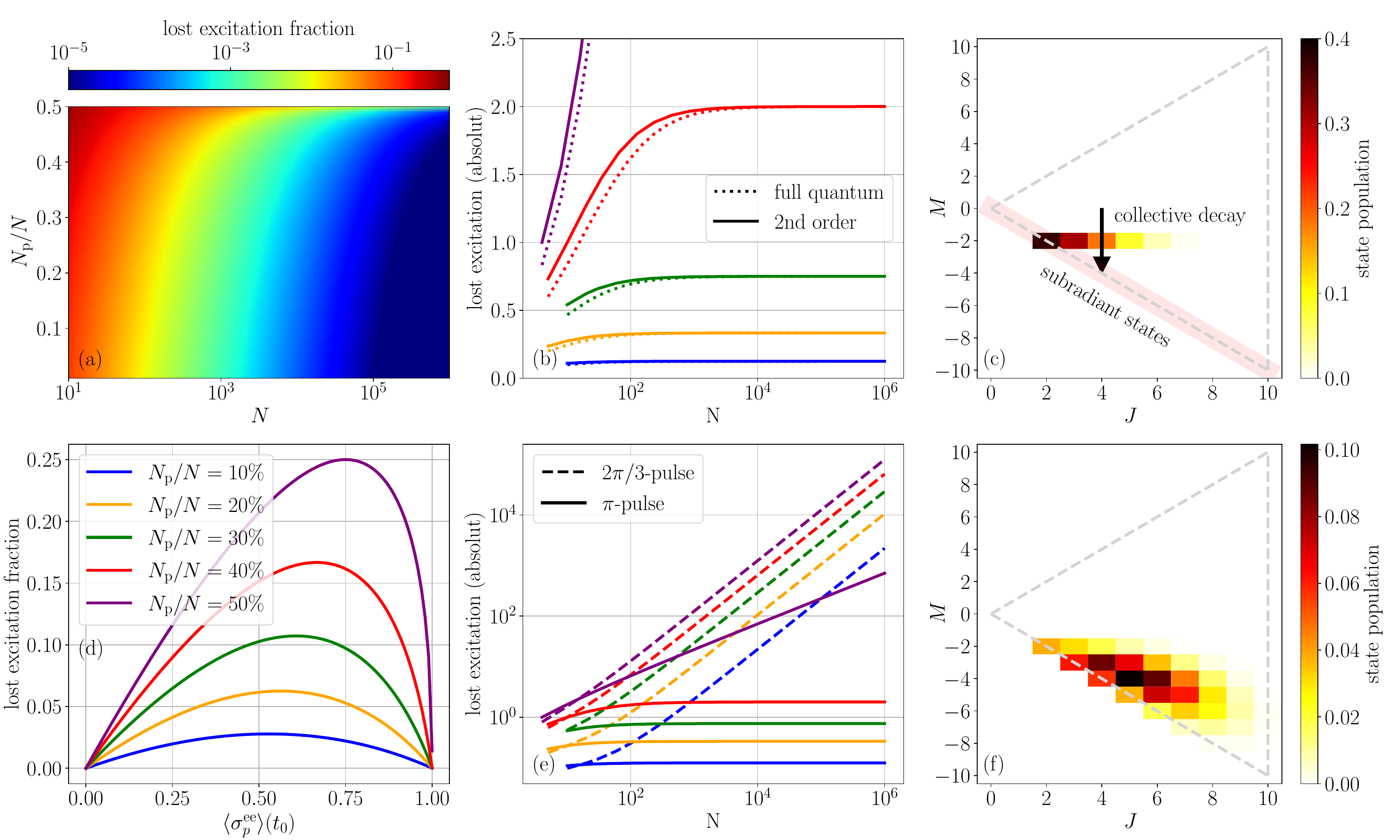}
    \caption{\emph{Subradiant states.} (a) Percentage of lost excitation for a full excitation of the number of pumped atoms $N_\mathrm{p}$. For large $N$ the fraction of lost excitation approaches zero. (b) Absolute number of lost excitations for different fractions of pumped atoms $N_\mathrm{p} / N$. Below the threshold, the number of lost excitations saturates to $N_\mathrm{p}/(N_\mathrm{np} - N_\mathrm{p})$. (d) Lost excitation fraction for coherent excitations to an initial excited state population of $\langle \sigma_\mathrm{p}^{ee} \rangle (t_0)$  for $N = 10^4$ atoms. Weaker coherent excitations can lead to more losses. (e) Lost excitation for an initial $2/3 \pi$-pulse. It does not saturate for large $N$. (c) and (f) Population of the Dicke states for $N = 20$ atoms and an initial $\pi$-pulse and $2/3 \pi$-pulse, respectively on $N_\mathrm{p} = 8$ atoms. In (c) the subradiant states (shaded red) and the collective decay (merely reducing $M$) are indicated.}
    \label{fig:lost_excitation_scans}
\end{figure*}
Subradiant states with respect to a single mode waveguide are of interest for various quantum technologies, in particular for controlled photon storage and release~\cite{asenjo2017exponential, holzinger2022control}. However, deterministically and reliably creating such states is experimentally difficult. In ref.~\cite{holzinger2022control} the authors showed that quasi-localized dark states can be created by exciting a minority fraction of the emitters separated by multiples of half an effective wavelength $\lambda_\mathrm{eff}/2$. In this case, the coherent coupling between the emitters vanishes ($\Omega_{ij}^\mathrm{1D} = 0$), and the dynamics reduces to waveguide-mediated collective emission. The mere presence of the coupled emitters in the ground state inhibits the superradiant decay of the excited fraction. So far only systems with a moderate number of emitters where a small fraction is fully excited have been considered. With a larger fraction of excited emitters the prepared states lose a certain amount of their excitation energy before the system becomes fully dark.

In the following, we analyze the corresponding dynamics in the limit of large numbers of emitters and investigate the scaling of the lost excitation fraction in the preparation process of these subradiant states via transverse coherent pumping. We start by assuming a spacing of $\lambda_\mathrm{eff}$ between the emitters. The discrete translation symmetry of the model allows us to describe it with two collective spins only, a spin-$N_\mathrm{p}/2$ for the pumped atoms and a spin-$N_\mathrm{np}/2$ for the non-pumped atoms. In fig.~\ref{fig:lost_excitation_scans}a we show the percentage of lost excitation depending on the total number of emitters $N = N_\mathrm{p} + N_\mathrm{np}$ coupled to the waveguide and the fraction of fully excited atoms $N_\mathrm{p} / N$. We observe that this percentage decreases with $N$. In particular, the suppression with growing $N$ dominates over the increased losses for larger fractional excitations. This means, that the percentage of lost excitations can be made arbitrarily small for a sufficiently large number of emitters, up to the threshold for subradiance at $N_\mathrm{p}/N = 50 \%$.

The number of lost excitations and the subradiance threshold can be nicely understood by means of the Dicke states $| J, M \rangle$~\cite{dicke1954coherence}, the eigenstates of the collective spin operators $J^z | J, M \rangle = M | J, M \rangle$ and $\mathbf{J}^2 | J, M \rangle = J(J+1) | J, M \rangle$, with $0 \leq J \leq N/2$ and $|M| \leq J$.  Figure~\ref{fig:lost_excitation_scans}c depicts the Dicke state representation for a full excitation of $N_\mathrm{p} = 8$ atoms and $N = N_\mathrm{np} + N_\mathrm{p} = 20$. We obtain a distribution of Dicke states with a well-defined $M$ quantum number corresponding to $N_\mathrm{p}$ excitations. Since the collective decay of a Dicke state appears as a reduction of $M$ with $J$ being unchanged ($J^- | J, M \rangle \rightarrow | J, M-1 \rangle$), one finds that all states with $M = -J$ are fully subradiant. It is also clear that only subradiant states with $M \leq 0$ can exist, corresponding to $N_\mathrm{p} \leq N/2$~\cite{shammah2018open, shankar2021subradiant, holzinger2022control, hotter2023cavity, bohr2024collective}. The average lost excitations then correspond to the population of the state $| \psi \rangle$ times its excitation difference to the subradiant state with the same $J$ quantum number to which it will decay, i.e.\ $\sum_{J,M} | \langle \psi | J,M \rangle |^2 (M + J)$. In fig.~\ref{fig:lost_excitation_scans}b we show the total number of lost excitations for different fractions of pumped atoms $N_\mathrm{p}/N$. We find that it converges to $N_\mathrm{p} / (N_\mathrm{np} - N_\mathrm{p})$ for large $N$. The dotted lines correspond to calculations using the Dicke states, which perfectly agree with full quantum simulations (not shown). The solid lines are the results of the second-order cumulant expansions, which agree very well with the full quantum model, especially for large $N$.

\begin{figure*}[tb]
    \centering
    \includegraphics[width=0.98\linewidth]{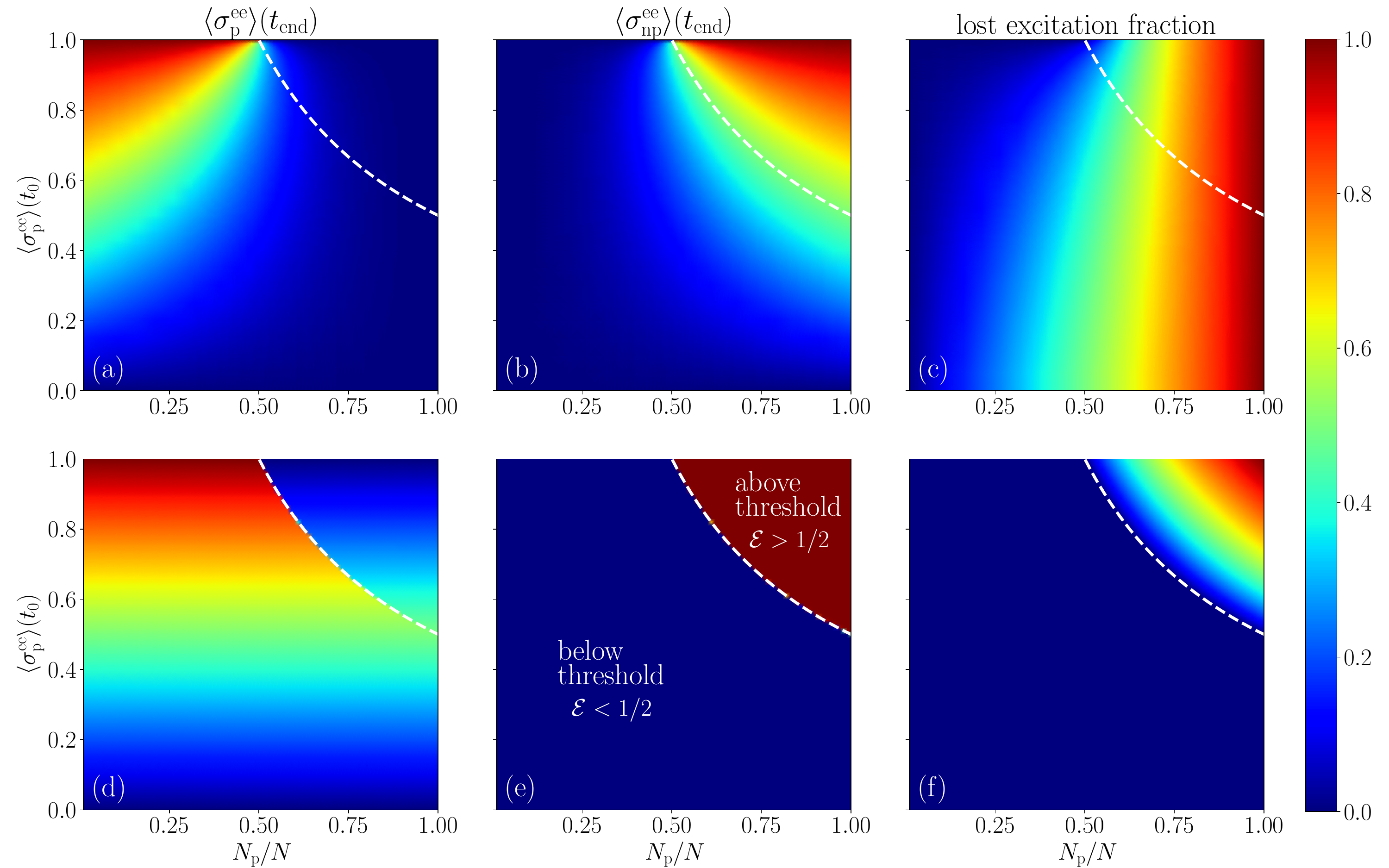}
    \caption{\emph{Steady-State Population and Lost Excitation.} Scan over the fraction of pumped atoms $N_\mathrm{p} / N$ and the initial excited state population $\langle \sigma^{ee}_\mathrm{p} \rangle (t_0)$ after a coherent excitation pulse. The upper panel depicts the case for an emitter spacing of $d = \lambda_\mathrm{eff}$ ($\Gamma^\mathrm{1D}_{ij} = \Gamma^\mathrm{1D}$, symmetric decay rates) and the lower the one for $d = \lambda_\mathrm{eff}/2$ ($\Gamma^\mathrm{1D}_{ij} = \pm \Gamma^\mathrm{1D}$, anti-symmetric decay rates). (a) and (d) show the steady-state population of the pumped atoms, (b) and (e) of the non-pumped atoms while (c) and (f) depict the fraction of lost excitations. The subradiant behavior and the energy transfer drastically improve for the anti-symmetric decay rates. For an initial full excitation, the behavior is the same for both cases. For the lower panel a threshold at $\mathcal{E} = 1/2$ (dashed white line) from subradiance to superradiance with subsequent superabsorption can be observed prominently. The chosen atom number in these scans is $N = 10^6$.}
    \label{fig:scan_2D_pm}
\end{figure*}

In fig.~\ref{fig:lost_excitation_scans}f we show the population of the Dicke states for an initial coherent excitation of $N_\mathrm{p} = 8$ atoms with a $2/3 \pi$-pulse ($\langle \sigma_\mathrm{p}^{ee} \rangle (t_0) = 75 \%$) and $N = 20$. Surprisingly, this state loses more excitations than with a $\pi$-pulse, although it is initially less excited. This is shown in Figure~\ref{fig:lost_excitation_scans}d and Figure~\ref{fig:scan_2D_pm}c for different initial coherent excitations and the number of pumped emitters. Figure~\ref{fig:lost_excitation_scans}e depicts the lost excitation as a function of the atom number $N$ for an initial $2/3 \pi$-pulse. In this case, it does not saturate for large $N$. This means, for a spacing between the emitters in multiples of $d = \lambda_\mathrm{eff}$ this dark state preparation protocol works best when a fraction of the emitters are fully excited. For a coherent excitation, this requires an almost perfect $\pi$-pulse.

Another way of creating subradiant states is to use opposite phases such that the emitted radiation interferes destructively~\cite{ballantine2020subradiance,rubies2022superradiance,hotter2023cavity, yan2023superradiant, bohr2024collective}. In a waveguide, this can be realized by a spacing of the atoms in multiples of $d = \lambda_\mathrm{eff}/2$.  In fig.~\ref{fig:scan_2D_pm}f we plot the percentage of lost excitation for this case. One can see that the lost excitation depends mainly on the initial excitation $\mathcal{E} = \langle \sigma^{ee}_\mathrm{p} \rangle (t_0) \cdot N_\mathrm{p} / N$. Below the threshold at $\mathcal{E} = 1/2$ the lost excitation fraction is very close to zero for a sufficiently large $N$, i.e.\ we can create subradiant states through an arbitrary coherent excitation almost without losing energy (for $\mathcal{E} < 1/2$).

\section{Superradiance and Superabsortion}
\begin{figure*}[tb]
    \centering
    \includegraphics[width=0.99\linewidth]{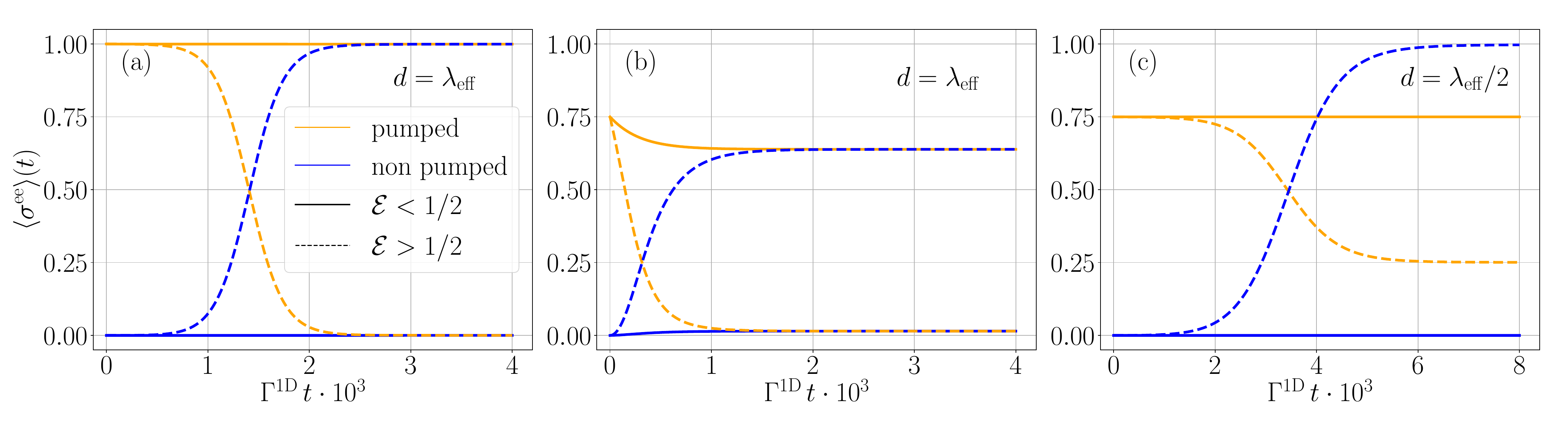}
    \caption{\emph{Time Evolution.} (a) Subradiant behavior (solid lines) and superradiant energy transfer (dashed lines) for initially fully inverted pumped atoms. (b) Coherent excitation to $75\%$ ($2/3 \pi$-pulse) for symmetric decay rates ($d = \lambda_\mathrm{eff}$), where subradiance and energy transfer are less optimal. (c) Anti-symmetric decay rates ($d = \lambda_\mathrm{eff}/2$), featuring subradiance and the optimal energy transfer for arbitrary coherent excitations. The chosen atom number is $N = 10^4$, for $\mathcal{E} < 1/2$ ($\mathcal{E} > 1/2$) we use $N_\mathrm{p} = 2 \cdot 10^3$ ($N_\mathrm{p} = 8 \cdot 10^3$).}
    \label{fig:example_timeevo}
\end{figure*}

In the previous section, we have focused on the collective subradiant behavior of the emitters coupled to a waveguide. From the Dicke triangle representation, it is clear that such subradiant states can only exist for a total number of excitations below $N/2$ (corresponding to $M \leq 0$). In the following, we investigate the dynamics of total initial excitations above $50\%$. As before, we start by considering a spacing of multiples of $\lambda_\mathrm{eff}$. In this case, we observe a fast excitation transfer from the initially fully excited to the non-excited ensemble facilitated by the collective decay via the shared bath, i.e.\ the waveguide~\cite{hama2018relaxation, hama2018negative, dias2023entanglement}. In fig.~\ref{fig:example_timeevo}a, we show a typical time evolution above the threshold ($\mathcal{E} > 1/2$, dashed lines), where we observe a superradiant decay of the initially excited ensemble and superabsorption in the initially non-excited ensemble. For comparison we have also plotted the behavior below the threshold ($\mathcal{E} < 1/2$, solid lines), where no energy transfer takes place.
 
Superabsorption can be interpreted as the inverse process to superradiance~\cite{higgins2014superabsorption, yang2021realization, quach2022superabsorption}. Hence, the excitation is usually immediately reemitted after the rapid collective absorption, which makes it difficult to observe the phenomenon. In the two-ensemble case described above, however, the emitters are in a subradiant state after the absorption, which means it is not followed by superradiant emission. This should make it experimentally much more accessible and useful. Let us mention that in the effective waveguide description the term superabsorption refers to the fast absorption of virtual photons.

Similar to the subradiance described in the previous section we find that the superradiance with subsequent superabsorption in the case of a spacing of multiples of $\lambda_\mathrm{eff}$ works well only if one ensemble is initially fully excited. Figure~\ref{fig:example_timeevo}b shows the dynamics for a coherent excitation of the pumped ensemble to $\langle \sigma_\mathrm{p}^\mathrm{ee} \rangle (t_0) = 75\%$ ($2/3 \pi$-pulse), where we notice that the energy transfer is not fully exciting the non-pumped ensemble above threshold ($\mathcal{E} > 1/2$, dashed lines). This can also be seen in the Scan~\ref{fig:scan_2D_pm}b.

Again, in order to optimize this process such that it works well for arbitrary coherent excitations, we can use an emitter spacing of multiples of $\lambda_\mathrm{eff}/2$. The resulting opposite phases in the collective decay lead to a sharp threshold at an initial excitation fraction of $\mathcal{E} = 1/2$. Below the threshold the system is subradiant and above we observe optimal superradiance from the pumped ensemble with subsequent superabsorption from the non-pumped ensemble. This is demonstrated in the time evolution in fig.~\ref{fig:example_timeevo}c and scans in fig.~\ref{fig:scan_2D_pm}d-f, the dashed line represents the threshold at $50\%$ total excitation.

Comparing the results for $\lambda_\mathrm{eff}$ and $\lambda_\mathrm{eff}/2$ spacing in fig.~\ref{fig:scan_2D_pm}a-c and fig.~\ref{fig:scan_2D_pm}d-f, respectively, we observe a profound difference in the threshold behavior for not completely inverted coherently excited ensembles.

\begin{figure}[tb]
    \centering
    \includegraphics[width=0.99\linewidth]{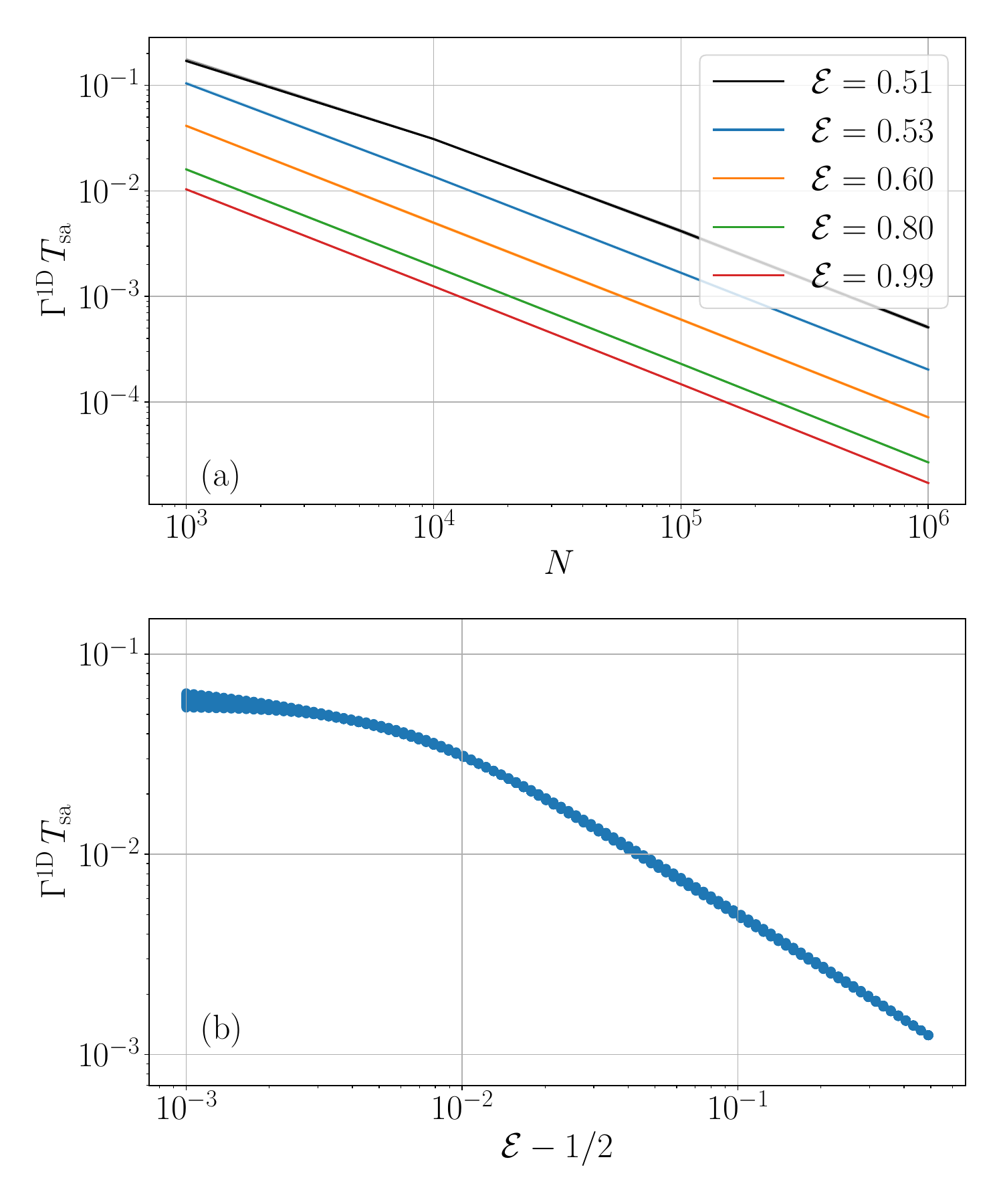}
    \caption{\emph{Superabsorption Time Scale.} (a) Atom number scaling of the energy transfer time $T_\mathrm{sa}$ as approximately $N^{-1}$ (b) Transfer time $T_\mathrm{sa}$ for different coherent excitations with the same initial excitation fraction $\mathcal{E} = \langle \sigma_\mathrm{p}^{ee} \rangle (t_0) N_\mathrm{p} / N $. Far above the threshold $T_\mathrm{sa}$ depends on $\mathcal{E}$ only. Close to the threshold, a full excitation of fewer emitters leads to a faster transfer than a smaller coherent excitation of more emitters. The number of atoms in (b) is $N = 10^4$.}
    \label{fig:timescale_sa}
\end{figure}

In order to characterize the superradiant energy transfer  fig.~\ref{fig:timescale_sa}a depicts the atom number dependence of the time $T_\mathrm{sa}$ it takes the non-pumped ensemble to reach $95\%$ of the final state excitation, i.e.\ $\langle \sigma^{ee}_\mathrm{np} \rangle = 0.95 \cdot \langle \sigma^{ee}_\mathrm{np} \rangle(t_\mathrm{end})$. We find a scaling close to $T_\mathrm{sa} \sim N^{-1}$ as for superradiant decay~\cite{dicke1954coherence, norcia2016superradiance, laske2019pulse, pennetta2022collective}. For an initial excitation sufficiently above the threshold this scaling holds for any coherent excitation. Close to the threshold different coherent excitations with the same initial excitation fraction $\mathcal{E}$ can lead to a slightly different transfer duration $T_\mathrm{sa}$. This can be seen in fig.~\ref{fig:timescale_sa}b, where we plot $T_\mathrm{sa}(\mathcal{E})$ for different coherent excitations.  In the vicinity of the threshold, the transfer is faster for a full excitation of fewer emitters (data points with smaller $T_\mathrm{sa}$) than a smaller coherent excitation of more emitters with the same $\mathcal{E}$.

\section{Uniform Distribution: Coherent Dipole-Dipole Interaction Effects}
\begin{figure*}[tb]
    \centering
    \includegraphics[width=0.99\linewidth]{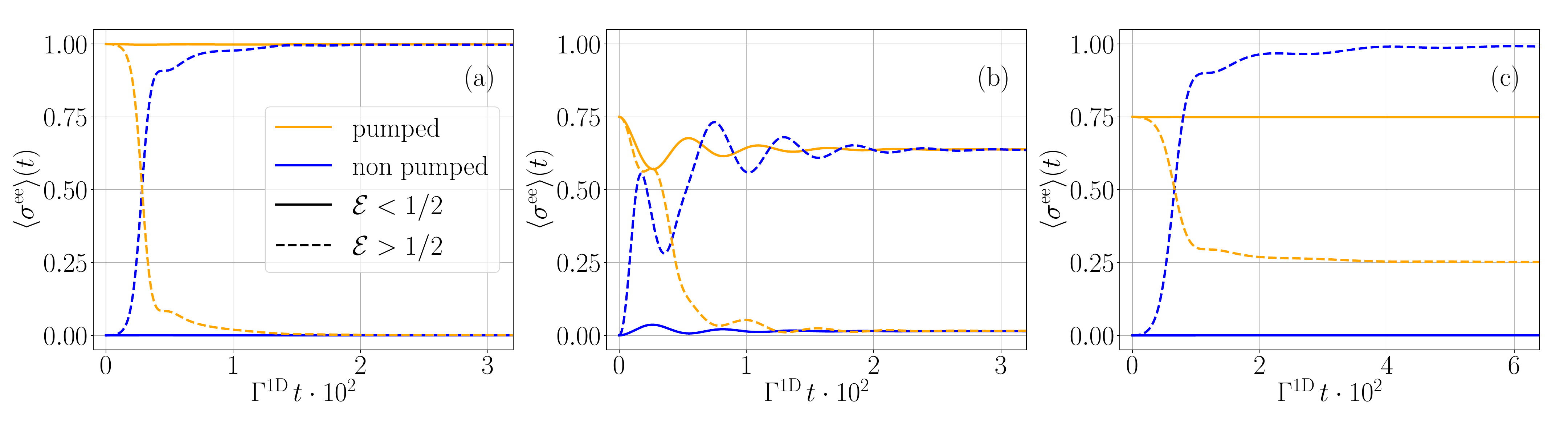}
    \caption{\emph{Time Evolution including Coherent Dipole-Dipole Interaction.} Population of the pumped and non-pumped emitters. For an initial full excitation of the pumped ensemble (a), the dynamics are only slightly modified by the coherent interaction. In contrast the behavior for a coherent excitation to $75\%$ changes significantly due to the coherent interaction (b). (c) shows that the optimal behavior can be recovered with a vanishing excitation phase in the pumped ensemble for coherent excitations. The atom numbers are the same as in Figure~\ref{fig:example_timeevo}.}
    \label{fig:example_timeevo_coh}
\end{figure*}

So far we have considered the special cases in which the emitters are located precisely at multiples of $\lambda_\mathrm{eff}/2$, where the coherent dipole-dipole interaction vanishes, i.e.\ $\Omega_{ij} = 0 $ for all $i$ and $j$. For a moderate number of emitters that can be positioned e.g.\ with optical tweezers or in the case of microwave waveguides this is experimentally feasible. However, for large atomic clouds that couple to an optical waveguide, all positions along the waveguide with respect to the periodic coupling are equally likely. In order to investigate the influence of the resulting coherent dipole-dipole interaction we sample the different positions. In particular, we use four equidistant positions with a spacing of $d = \lambda_\mathrm{eff}/4$ for the pumped and non-pumped ensemble. A comparison with six equidistant positions is shown in fig.~\ref{fig:coh_4x4_6x6_all_populations}a, b, d and e in the appendix, which underpins that four ensembles are sufficient for a qualitative description of the coherent dipole-dipole interaction effects.

In fig.~\ref{fig:example_timeevo_coh} we show the dynamics for $N = 10^4$ emitters for different cases. For a fully excited pumped ensemble (Figure~\ref{fig:example_timeevo_coh}a) the behavior below as well as above threshold is only weakly modified by the coherent dipole-dipole interaction. The slower overall dynamic mainly stems from the reduced incoherent rates. This has been tested by simulating the dynamics with incoherent rates $\Gamma_{ij}$ corresponding to a spacing $d = \lambda_\mathrm{eff}/4$ while neglecting the coherent interaction ($\Omega_{ij} = 0$), see fig.~\ref{fig:coh_4x4_6x6_all_populations}c and f in the appendix. 
In contrast, for an arbitrary coherent excitation, the subradiant and superabsorption features are strongly affected by the coherent dipole-dipole interaction. The dynamics for an initial coherent $2/3 \pi$-pulse is shown in fig.~\ref{fig:example_timeevo_coh}b. We see that the behavior is not optimal, similar to the one with a spacing of $d = \lambda_\mathrm{eff}$ (Figure~\ref{fig:example_timeevo}). The coherent interaction seems to cancel the positive effect from the opposite phases of the collective decay rates.

The crucial difference in the above two cases is the initial coherence in the later one. This means an initial incoherent excitation to $\langle \sigma_\mathrm{p}^{ee} \rangle (t_0) = 75 \%$ should lead to the same optimal subradiant and superradiant transfer behavior. One way to approximately resemble an incoherent excitation is to excite ensembles with vanishing phases. In fig.~\ref{fig:example_timeevo_coh}c we show the dynamics for this case, where the pumped ensemble is divided into two sub-ensembles which are excited with opposite phases. This leads to an almost optimal behavior below and above the threshold. Note that for a large ensemble of emitters an overall vanishing excitation phase is usually automatically given due to the position distribution. These results suggest that the phenomenon should be observable with two separated dilute clouds of atoms coupled to a waveguide.

\section{Conclusions}
We have shown a transition from sub- to superradiance with subsequent superabsorption for two ensembles of emitters coupled to a waveguide. The lost excitation for the subradiant states saturates for a large number of emitters. This allows for creating highly excited subradiant states with up to $50\%$ excitations. Above the threshold an efficient superradiant energy transfer between the two ensembles takes place. We have demonstrated that for a not fully inverting coherent excitation of the pumped atoms, a vanishing phase is crucial for optimal behavior. This is, however, usually automatically given for large atomic ensembles. The necessary operating conditions are within reach of current experimental setups~\cite{okaba2019superradiance,abend2023technology,bertoldi2021fast,melli2024azimuthal}.

The efficiency of the process can be increased by optimizing the emitter positions, this can e.g.\ be realized with optical tweezers~\cite{pesce2020optical} or self-organization~\cite{chang2013self, ritsch2013cold}. The described mechanisms should also be valid for multi-mode waveguides. In future work, it could be interesting to investigate potential retardation effects for widely separated ensembles, as well as the possibility of engineering the dissipation by detuning ensembles. We also want to mention that ensembles of emitters in a single-mode bad cavity, where the cavity field can be adiabatically eliminated, feature a similar collective dissipation with opposite phases. Yet, for a vanishing atom-cavity detuning the resulting effective equations for the atoms do not experience coherent dipole-dipole interaction terms~\cite{gardiner2004quantum, shankar2021subradiant, jaeger2022lindblad}. This can be beneficial for the above-described phenomena. Of particular interest are ring cavities with a uniform coupling strength for all atoms~\cite{zhang2023development}.

\section*{Acknowledgments}
The numerical simulations were performed with the open-source frameworks QuantumCumulants.jl~\cite{plankensteiner2021quantumcumulantsjl} and  QuantumOptics.jl~\cite{kramer2018quantumoptics}. M.\ F.\ and H.\ R.\ acknowledge funding from the Austrian Science Fund (FWF) doctoral college DK-ALM W1259-N27.  M.\ F.,\  L.\ O.\ and H.\ R.\ acknowledge the FET OPEN Network Cryst3 funded by the European Union (EU) via Horizon 2020.

\bibliography{references}

\clearpage
\widetext

\section*{Appendix}
\subsection{Simulation Code}
The Julia code to derive and numerically solve the second-order cumulant equations can be found in the online documentation of the toolbox QuantumCumulants.jl~\cite{plankensteiner2021quantumcumulantsjl}: \\
\url{https://qojulia.github.io/QuantumCumulants.jl/stable/examples/waveguide/}. By setting $M_\mathrm{p} = 4$ and $M_\mathrm{np} = 4$ in the example one creates the 324 equations to reproduce the results in Figure~\ref{fig:example_timeevo_coh}a and b.

\subsection{Six vs. Four Subensembles}
\begin{figure*}[h]
    \centering
    \includegraphics[width=0.99\linewidth]{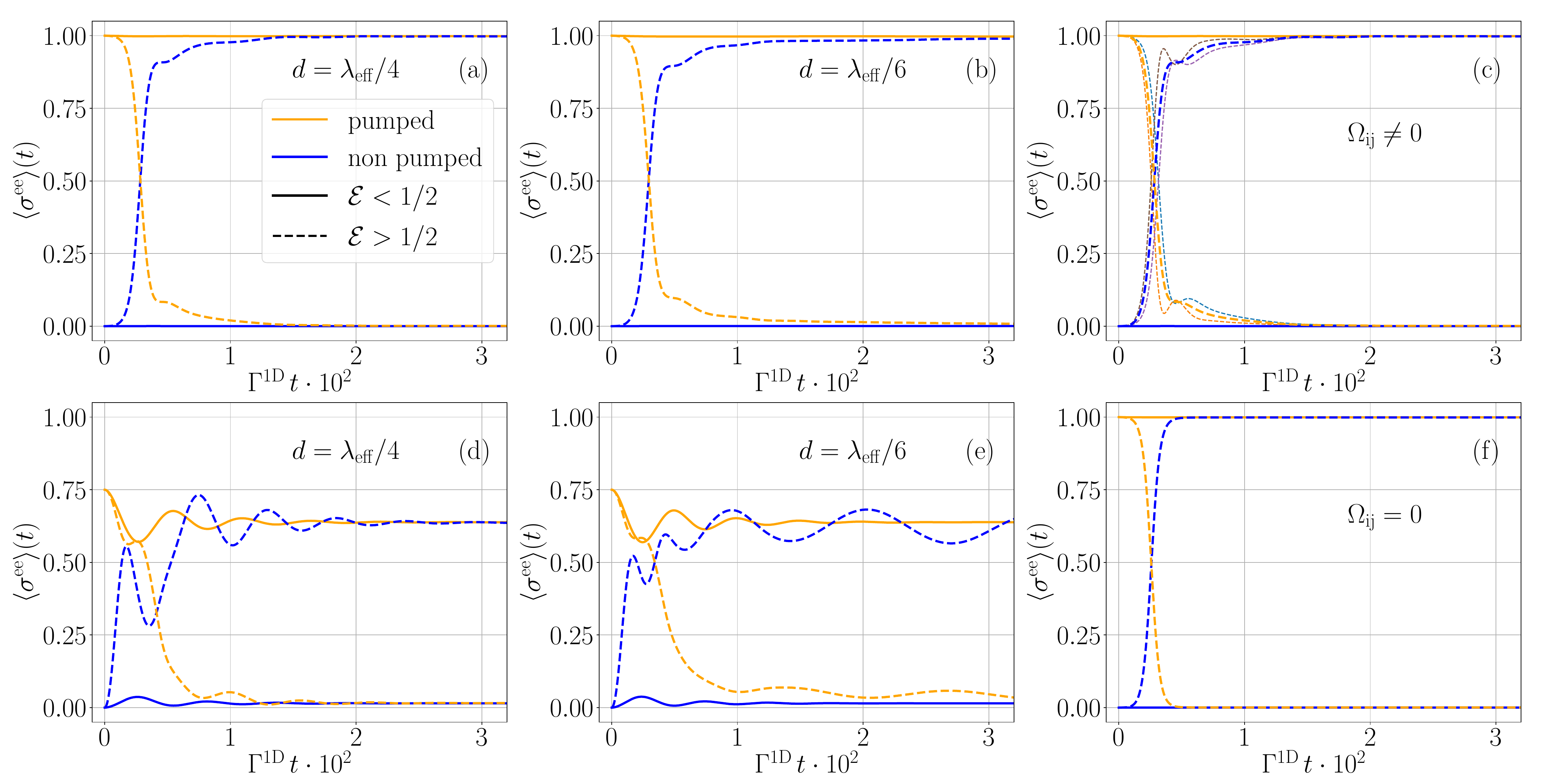}
    \caption{\emph{Coherent Dipole-Dipole Interaction Effects.} (a) and (d) Time evolution above and below the threshold for an initial $\pi$-pulse and $2/3 \pi$-pulse excitation, respectively. These plots are the same as in Figure~\ref{fig:example_timeevo_coh} for four positions with a spacing of $d = \lambda_\mathrm{eff}/4$. (b) and (e) Comparison to six sub-ensembles with a spacing of $d = \lambda_\mathrm{eff}$/6. The time evolution in (a) and (b) is almost identical, for (d) and (e) they agree qualitatively. In (c) we plot all trajectories of the different ensembles for the case presented in (a). In (f) we artificially set the coherent dipole-dipole interaction to zero. This reveals that the slower dynamic for the four positions can be traced back to the reduced collective decay rate.}
    \label{fig:coh_4x4_6x6_all_populations}
\end{figure*}

In fig.~\ref{fig:coh_4x4_6x6_all_populations} we compare the time evolution including coherent dipole-dipole interaction for four and six different positions at $x_4 = \{ 0, 1, 2, 3 \} \cdot \lambda_\mathrm{eff}/4$ and $x_6 = \{ 0, 1, 2, 3, 4, 5 \} \cdot \lambda_\mathrm{eff}/6$, respectively. For the full excitation in fig.~\ref{fig:coh_4x4_6x6_all_populations}a and b the difference is barely visible. For an initial coherent excitation to $75\%$ the oscillations are stronger, see Figure~\ref{fig:coh_4x4_6x6_all_populations}d and e. However, we expect the oscillations to damp out very quickly for more positions. Unfortunately, we cannot test this in the second-order cumulant expansion approach, since the number of coupled differential equations for the $2 \times 6$ spins is already $702$. In fig.~\ref{fig:coh_4x4_6x6_all_populations}c we show the dynamics for each of the four ensembles. In fig.~\ref{fig:coh_4x4_6x6_all_populations}f we set the coherent dipole-dipole interaction to zero, and see that this only slightly changes the dynamics.

\end{document}